\documentclass[]{aa}  

\usepackage{graphicx}
\usepackage{txfonts}
\usepackage{booktabs}
\usepackage{natbib}
\usepackage{hyperref}
\usepackage{footnote}
\usepackage{tablefootnote}
\usepackage{longtable}
\usepackage{url}
\usepackage{color}
\usepackage{float}
\usepackage{rotating}
\usepackage{multirow}
\usepackage{xcolor}

\newcommand{\vmic}{\xi_{\rm t}}
\newcommand{\Vmic}{\xi_{\rm t}}
\newcommand{\vmac}{V_{\rm mac}}

\newcommand{\mA}{{\rm m\AA}}
\newcommand{\Elow}{E_{\rm low}}
\newcommand{\Eup}{E_{\rm up}}

\newcommand{\Eu}[5]{\mbox{$#1\,^#2{\rm #3}^{{\rm #4}}_{\rm #5}$}}
\newcommand{\Y}[5]{\mbox{$#1\,^#2{\rm #3}^{{\rm #4}}_{\rm #5}$}}
\begin{document}

\title{3D NLTE modelling of Y and Eu}
\subtitle{Centre-to-limb variation and solar abundances}
\author{N. Storm\inst{1}
\and
P. S. Barklem \inst{2}
\and
S. A. Yakovleva \inst{3}
\and
A. K. Belyaev \inst{3}
\and
P. Palmeri \inst{4}
\and
P. Quinet \inst{4,5}
\and
K. Lodders \inst{6}
\and
M. Bergemann\inst{1}
\and
{R. Hoppe\inst{1}}
}

   \institute{$^1$ Max-Planck Institute for Astronomy, K\"{o}nigstuhl 17, D-69117 Heidelberg, Germany\\
   $^2$ Theoretical Astrophysics, Department of Physics and Astronomy, Uppsala University, Box 516, 751 20 Uppsala, Sweden\\
   $^3$ Institute of Physics, Herzen University, 191186 St. Petersburg, Russia\\
   $^4$ Physique Atomique et Astrophysique, Universit\'e de Mons - UMONS, B-7000 Mons, Belgium\\
   $^5$ IPNAS, Universit\'e de Li\`ege, B-4000 Li\`ege, Belgium\\
   $^6$ Planetary Chemistry Laboratory, Dept of Earth \& Planetary Sciences and McDonnell
Center for Space Sciences, Washington Univ, St Louis MO, USA\\
   }

   \date{Received XX XX XXXX; accepted XX XX XXXX}

% \abstract{}{}{}{}{} 
% 5 {} token are mandatory
 
  \abstract
   {Abundances of s- and r-process elements in Sun-like stars constrain nucleosynthesis in extreme astrophysical events, such as compact binary mergers and explosions of highly magnetised rapidly rotating massive stars.}
   {We measure solar abundances of yttrium (Y) and europium (Eu) using 3D non-local thermal equilibrium (NLTE) models. We use the model to determine the abundance of Y, and also explore the model's ability to reproduce the solar centre-to-limb variation of its lines. In addition, we determine the Eu abundance using solar disc-centre and integrated flux spectra.}
   {We developed an NLTE model of Eu and updated our model of Y with collisional data from detailed quantum-mechanical calculations. We used the IAG spatially resolved  high-resolution
solar spectra to derive the solar abundances of Y across the solar disc and of Eu for integrated flux and at disc {centre} using a set of carefully selected lines and a 3D radiation-hydrodynamics model of the solar atmosphere.}
   {We find 3D NLTE solar abundances of {A(Y)$_{\textrm{3D NLTE}}~$~=~$2.30 \pm 0.03_{\textrm{stat}} \pm 0.07_{\textrm{syst}}$} dex based on observations at all angles and {A(Eu)~=~$0.57 \pm 0.01_{\textrm{stat}} \pm 0.06_{\textrm{syst}}$} dex based on the integrated flux and disc-centre intensity. {3D} NLTE modelling offers the most consistent abundances across the solar disc, and resolves the problem of severe systematic bias in Y and Eu abundances inherent to 1D LTE, 1D NLTE, and 3D LTE modelling.}
   {}
   \keywords{ Atomic data -- Sun: abundances -- Sun: atmosphere -- Line: formation --  Methods: observational}
   \maketitle
%
%--------------------------------------------------------------------

\section{Introduction}
\label{sec:intro}

Nuclear neutron-capture processes are responsible for the synthesis of approximately two-thirds of the elements in the periodic table; specifically, those heavier than the iron peak elements. They are classified into three groups, namely slow (s)-, intermediate (i)-, and rapid (r)-processes, depending on the timescale differences of $\beta$-decay and neutron capture, and on the flux of neutrons available to the system in given astrophysical conditions. S-process elements are thought to be mainly produced in intermediate-mass stars during the asymptotic giant branch (AGB) phase \citep[e.g.][]{Busso1999, Cristallo2011, Karakas2016}. R-process sites are heavily debated; these include neutron-driven winds in core-collapse SNe \citep[e.g.][]{Takahashi1994, Woosley1994, Arcones2013, Bliss2018}, explosions of rapidly rotating magnetised massive stars \citep[e.g.][]{Siegel2017, Halevi2018, Siegel2019, Reichert2023}, and neutron-star mergers \citep[e.g.][]{Rosswog1999, Halevi2018, Siegel2019, Watson2019}. Finally, i-process sources are considered to be post-AGB stars \citep{Herwig2011}, super-AGB stars \citep{Jones2016}, low-metallicity AGB stars \citep{Karinkuzhi2021}, and accreting white dwarfs \citep{Denissenkov2019}. Accurate abundance determination of neutron-capture elements in cool stars, such as the Sun, are therefore important for constraining the nucleosynthetic production sites. The abundances of these elements in the Sun represent the baseline for spectroscopic studies of all other stars.

Yttrium (Y) is an odd-Z first-peak s-process element \citep{Arlandini1999, Bisterzo2014, Kobayashi2020}. However, all earlier estimates of the solar Y abundance relied on either 1D local thermodynamic equilibrium (LTE), 3D LTE, or 1D non-local thermodynamic equilibrium (NLTE) modelling. Solar Y estimates include those of \citet{Hannaford1982}, A(Y)\footnote{We use the standard notation of A(X) to represent the abundance of the element X relative to hydrogen, where the quantity A is defined as log$_{10}$ ($N_{\rm X}/N_{\rm H})+12$, whereas $N_{\rm X}$ is the number of atoms per unit volume.} $= 2.24$, \citet{Grevesse2015} in 3D LTE, A(Y) = $2.21$, and our recent analysis in 1D NLTE, which gave A(Y)$_\textrm{NLTE}$ $ = 2.12$ \citep{Storm2023}. The Y abundance in CI chondrites is estimated at $2.17 \pm 0.04$ dex \citep{Lodders2009}.  

Europium (Eu) is usually viewed as an almost pure r-process element \citep[e.g.][]{Otsuki2006, Cowan2021}. The solar europium abundance was derived in 1D NLTE by \citet{Mashonkina2000}, giving A(Eu) $= 0.53$, and in 3D by \citet{Grevesse2015}, giving $0.52 \pm 0.04$, with both estimates similar to the meteoric value of $0.51 \pm 0.02$ \citep{Lodders2009}. In \citet{Grevesse2015}, a flat $+0.03$ dex 1D NLTE correction was applied to the 3D LTE Eu abundances. This correction was estimated on the basis of calculations presented in \citet{Mashonkina2000}, where the model atom of Eu employed the Drawin recipe to describe collision processes of Eu with H atoms. No self-consistent 3D NLTE estimate of the solar Eu has been carried out to date.

We developed a comprehensive NLTE model of Eu and updated our model of Y with quantum-mechanical data for inelastic processes in collisions with hydrogen, and analysed the solar abundances of these two elements in full 3D NLTE. For Y, we explored the centre-to-limb variation (CLV), as done in other 3D NLTE studies \citep{Lind2017, Bergemann2021, Pietrow2023}, which provides a powerful test of line formation and consequently of the accuracy of the resulting abundance measurement \citep[e.g.][]{Holweger1978, LohnerBottcher2018}. However, for Eu, given the blend uncertainties, we opted to analyse solar disc-centre intensity and integrated flux instead. We present our analyses and the resultant findings in the present paper, which is organised as follows{: A description of} the details of the model atoms, the physical models, and measurement methods we used {are given} in Sect. \ref{sec:methods}{. The} solar abundances of Y and Eu, and {our analysis of the CLV variation of the diagnostic Y II lines} are presented in Sect. \ref{sec:results}{. Finally, our} results are briefly discussed and then summarised in Sect. \ref{sec:discussion}.
\section{Methods}\label{sec:methods}

\subsection{NLTE models}

The NLTE model of Y was adopted from \citet{Storm2023}, with modifications as described below. Briefly, the model comprises 187 and 236 fine-structure energy levels of Y I and Y II, with ionisation thresholds of $6.22$ eV and $12.23$ eV, respectively. The model is closed with the ground state of Y III. The basic energy level structure ---with fine-structure levels preserved--- and the radiative bound--bound transitions are adopted from the Kurucz database \citep{Kurucz1995}. Specifically, experimental values of energies for many levels are available from \citet{Nilsson1991}. In total, we include $11\,819$ radiative bound--bound transitions and 423 radiative bound--free transitions. Rate coefficients for the processes in the collisions with free electrons $e^{-}$  are estimated using the formulae from \citet{Regemorter1962} and \citet{Seaton1962}. 

In this work, we include new rate coefficients for the excitation and charge-transfer reactions in collisions with hydrogen atoms and negative ions. The data for inelastic processes are calculated for $107$ scattering channels of Y$^{+}$ + H and two ionic states of Y$^{2+}$ + H$^-$. In these calculations, we include states with experimentally derived asymptotic energy values and average over the total angular momentum quantum number {(J)}. Rate coefficients are calculated using the multichannel Landau-Zener asymptotic approach \citep{BelyaevPRA2013, YVB_AA_2016}. The calculated data are different from the data presented in \citet{Wang2023}, where the rate coefficients are calculated using the simplified quantum model that uses the two-state Landau-Zener model \citep{BY_AA_2017_1, BY_AA_2017_2}. We note that these datasets are available for J-averaged states, but our model atom explicitly includes fine structure. We therefore adopt the same rate coefficients for each fine structure state of a term. As shown by \citet{Bergemann2019}, applying more sophisticated re-distribution methods does not influence the statistical equilibrium. In total, $14\,517$ Y II transitions are now covered by data for H-induced excitation processes, and the charge-transfer rates are included 
for $199$ Y II states.

Here, we develop an NLTE model of Eu using the data from NIST \citep[original data from][]{Martin1978, Nakhate2000, Johnson2017} and Kurucz \footnote{\url{http://kurucz.harvard.edu/atoms/6301/}} databases. The electronic structure of Eu is represented by 662 levels in total, with Eu I represented by 498 levels and Eu II by 163 levels with the ionisation potentials of $5.67$ eV and $11.24$ eV, respectively. The model is closed by the lowest Eu III state, $4f$ $8^{}$S$^{\rm o}$. The energy levels and statistical weights were assembled from the NIST database, and only levels connected by radiative transitions tabulated in the Kurucz datasets were retained. We further applied a cut to the transition probabilities in order to eliminate extremely weak transitions with an $f$-value of less than $10^{-10}$ and limited the wavelength range up to $15$ $\mu$m. Fine structure was preserved for all states, where available. 

For collisional excitation and ionisation with $e^{-}$, we employed the same recipes as for Y. The excitation and charge-transfer processes involving H and H$^-$ collisions with Eu$^+$ and Eu$^{2+}$, respectively, were computed using the {linear combination of atomic orbitals (LCAO)} approach \citep{barklem_excitation_2016,barklem_erratum:_2017} for the electron-transfer mechanism, supplemented with the Kaulakys model \citep{Kaulakys1985, Kaulakys1991} for the momentum-transfer mechanism in Eu$^+$+H collisions.  This approach, namely of combining these two models with different physical mechanisms, was motivated by both empirical and theoretical findings \citep[see][]{amarsiInelasticCollisions7772018}. The LCAO model and codes have been extended {by the authors} to handle the case where the covalent states involve a singly ionised atom, and the ionic state a doubly ionised atom; this simply requires a change in the charge on the core of the target atom/ion X, and corresponding adjustments to the Hamiltonian, and therefore also to the relevant wavefunctions and matrix elements \citep[see][]{Barklem2016}. The Kaulakys code \citep{barklem_kaulakys_2015} was also adjusted by extending the code, which
previously only handled neutral atoms, to include momentum-space wavefunctions \citep{barklem_mswavef_2015}, so that it can work for ionised atoms \citep[see][Appendix 5]{Bransden2003}.

{Both of the Y and Eu model atoms, together with the previously published ones, are {available at} \footnote{\url{https://keeper.mpdl.mpg.de/d/6c2033ef5c5d4c9ca8d1/}}} in their respective folders.
\subsection{Diagnostic lines}
\label{subsec:diag_lines}

\begin{table}
\begin{minipage}{\linewidth}
\renewcommand{\footnoterule}{} 
\setlength{\tabcolsep}{3pt}
\caption{Diagnostic Y II and Eu II lines used for the solar-abundance analysis in this work. The $\log gf$ values are adopted from \citet[][their uncorrected values]{Palmeri2017} for Y and from \citet{Lawler2001} for Eu.}
\label{tab:y_atomic_data}     
\begin{center}
\begin{tabular}{l ccccc}
\noalign{\smallskip}\hline\noalign{\smallskip}  
$\lambda$ [\AA] & States & $\Elow$ & $\Eup$ &  $\log gf$ \\
 & & [eV] & [eV]  \\
\noalign{\smallskip}\hline\noalign{\smallskip}

  &  &  &  &  \\
Y II      &  &  &  & \\
4883.682 & \Y{a}{3}{F}{}{4} - \Y{z}{3}{D}{o}{3} & 1.08 & 3.62 & ~~0.15 \\
4900.118 & \Y{a}{3}{F}{}{3} - \Y{z}{3}{D}{o}{2} & 1.03 & 3.56 & $-$0.02 \\
5087.418 & \Y{a}{3}{F}{}{4} - \Y{z}{3}{F}{o}{4} & 1.08 & 3.52 & $-$0.18 \\
5200.406 & \Y{a}{3}{F}{}{2} - \Y{z}{3}{F}{o}{2} & 0.99 & 3.38 & $-$0.62 \\
Eu II      &  &  &  & \\
6645.104 & \Eu{a}{9}{D}{o}{6} - \Eu{z}{9}{P}{}{5} & 1.38 & 3.24 & 0.12 \\
\\
\noalign{\smallskip}\hline\noalign{\smallskip}
\end{tabular}
\end{center}
\end{minipage}
\end{table} 

To ensure the most robust solar abundance diagnostics, here we focus on the lines of Y and Eu that are sufficiently resolved and minimally affected by blends in the solar spectrum, and for which accurate experimental atomic data are available. Following our previous analysis in \citet{Storm2023}, we include four lines of Y II in the optical wavelength range, disregarding the redder features that are too weak for a reliable analysis. The atomic parameters of these lines are provided in Table \ref{tab:y_atomic_data}. Y is represented by a single stable isotope ($^{89}$Y) and the hyperfine splitting (HFS) is negligibly small \citep{Hannaford1982}. The transition probabilities for Y II lines were adopted from \citet{Palmeri2017}. We have opted to use the uncorrected values because the experimental lifetimes used in \citet{Palmeri2017} to correct the $f$-values for these transitions were measured by \citet{Hannaford1982} and by \citet{Wannstrom1988} and we suspect that they are slightly too long, as highlighted by \citet{Biemont2011}. Indeed, \citet{Biemont2011} stated that the shorter excitation laser pulses employed in their own measurements explain that they obtained systematically shorter and more accurate experimental values than the older ones published by \citet{Hannaford1982} and \citet{Wannstrom1988}. In addition, it was also shown by \citet{Biemont2011} that their {Hartree–Fock model with core-polarisation (HFR+CPOL)} calculations were in very good agreement (within the 10\% estimated error bars) with the {time-resolved laser-induced  fluorescence (TR-LIF)} experimental lifetimes. Therefore, we expect the uncorrected HFR+CPOL $f$-values presented in Table~\ref{tab:y_atomic_data} to be accurate up to $\sim$10\%.

The most reliable, albeit rather weak ($\sim$ 5 m$\AA$), Eu II line in the solar spectrum corresponds to the transition between the \Eu{a}{9}{D}{o}{6} and \Eu{z}{9}{P}{}{5} states at 6645.104 $\AA$.  Accurate laboratory $f$-values for Eu II were presented in \citet{Lawler2001}. These values were determined by combining the lifetimes measured by means of the {TR-LIF} technique and the branching fractions measured from Fourier transform spectroscopy. The uncertainty of the $f$-values was estimated at $12\%$. Following the methods described in \citet{Bergemann2010b}, we compute the HFS and isotopic structure for this line using the HFS magnetic dipole and electric quadrupole constants from \citet{Villemoes1992}. The solar isotopic abundance ratio for the two most abundant isotopes $^{151}$Eu and $^{153}$Eu is set to 47.81:52.19, respectively \citep{Lodders2009}. The individual HFS and isotopic components are co-added within $0.005$ $\mA$; therefore, in our model the 6645 \AA~line is represented by 11 hyperfine and isotopic components.
\subsection{Observations and spectrum synthesis}
\label{subsec:obs}

We employed the recently published solar intensity spectra \citep{Reiners2023,Ellwarth2023}, which were obtained with the Fourier Transform Spectrograph {(FTS)} mounted on the Vacuum Vertical Telescope at the Institut f\"{u}r Astrophysik G\"{o}ttingen (IAG). The spectra were taken at $14$ different angles $\mu$ ($\mu=\cos{\theta}$, where $\theta$ is the viewing angle with respect to the solar disc centre) from 1.00 to 0.20 and cover the wavelength range from  4200 to 8000 $\AA$, with the resolving power of $R = 700\,000$ at 6000 $\AA$. The signal-to-noise ratio (S/N) of the spectra varies from 420 to 690 per pixel. For comparison, we also use the  solar Kitt Peak National Observatory (KPNO) FTS atlas and intensity atlases \citep{Kurucz1984} with $R \approx 400\,000$.

All synthetic spectra were computed self-consistently using the {MULTI3D at Dispatch} code {\citep{Eitner2023}}
\footnote{\url{https://dispatch.readthedocs.io/en/latest/overview/index.html}}, which is a significantly updated version of the MULTI3D code \citep{Botnen1997, Leenaarts2009}. In this work, we use the 1D plane-parallel MARCS model \citep{Gustafsson2008} and the 3D radiation-hydrodynamics Stagger model \citep{Bergemann2012a, Magic2013}, and adopt the standard value of micro-turbulence $\vmic = 1$ km/s for the former \citep{Bergemann2012a}.

LTE and NLTE radiative transfer in 3D {were performed as explained} below. As in \citet{Bergemann2019} and \citet{Bergemann2021} (see also the series of tests in {\citet{Eitner2023}}), we explored a large range of spatial resolutions using 3D LTE calculations. NLTE statistical equilibrium was solved in the boxes with the resolution of (x,y,z) $=(30,30,230)$; however, the final analyses of Y and Eu abundances were carried out at the resolution equivalent to (x,y,z) $=(120,120,230)$. Intensities for the CLV work were computed at $13$ angles to encompass the range of observations provided by the IAG atlas.

The abundance analysis was carried out by fitting the line profiles directly to the data using the standard least-squares minimisation technique, as used in our previous 3D NLTE studies for example \citep{Bergemann2019, Bergemann2021}, using Scipy's curve fitting function \citep{2020SciPy-NMeth}. To account for blends, we also performed a careful spectrum synthesis of relevant transitions falling within the range relevant to the diagnostic Y II and Eu II lines. Specifically, we included H in 3D using the atomic model from \citet{Mashonkina2008}. However, we find that the abundance calculations for Y II are not sensitive to the detailed treatment of blends. Unfortunately, the Eu II 6645 \AA~line is significantly contaminated by a series of weak blends that cannot be described well using the standard line list from the Gaia-ESO survey \citep{Heiter2021}. Also, the VALD line list does not yield a good description of the corresponding wavelength range. This uncertainty is folded into the total error of the solar Eu abundance, as described below (Sect \ref{subsec:error_analysis}).
\subsection{Error analysis}
\label{subsec:error_analysis}
The analysis of statistical and systematic errors closely follows the methodology of \citet{Bergemann2012a}. As described in Sect. \ref{subsec:obs}, blends were computed self-consistently in 3D. We estimated the corresponding uncertainty in the abundance by examining its change between fitting with and without the other lines from the Gaia-ESO line list \citep{Heiter2021}. This resulted in an average blending error of $\sim 0.05$ dex for the Y II lines (line 4900 having the highest blending) and $\sim 0.03$ dex for the Eu II line. We also adopt the uncertainty of $\sim 0.04$ dex for Y II (assuming 10\% $f$-value uncertainty; see the discussion in Sect. \ref{subsec:diag_lines}) and $\sim 0.05$ dex for Eu II (based on the reported $\log gf$ uncertainty from \citet{Lawler2001}).

{The individual error components were calculated by adding the systematic uncertainty (including blends and uncertainty of $f$-values) and the fitting error in quadrature. For Y II, the statistical error was determined using the standard deviation among abundances derived from individual spectral lines. This approach was not applicable for Eu II, as only one line was fitted. Consequently, the statistical error for Eu II was determined by calculating the standard deviation from fits of different spectra. Overall, the $\log gf$ error constitutes the primary source of uncertainty in the final derived abundances.}
\section{Results}
\label{sec:results}

As we explore further in the following subsections, 3D NLTE analysis provides a more consistent abundance of chemical elements across the solar disc, particularly when compared to 1D LTE and 3D LTE approaches. Physically, this is due to two factors. First, the LTE lacks realistic treatment of the interaction between the plasma and radiation field, unlike {NLTE}. Secondly, 1D hydrostatic models lack realistic convection and turbulent velocity fields; this shortfall is usually mitigated by including scaling coefficients to the classical mixing-length theory of convection, such as convective efficiency scaling, \citep{Fuhrmann1993, Bernkopf1998, Grupp2004} and ad hoc broadening parameters \citep[e.g.][]{Jofre2019} to compensate for the missing physics. For example, one such adjustment was used by \citet[][]{Holweger1978} for example in the form of angle-dependent micro- and macro-turbulence. Clearly, such modifications only highlight the deficiencies of hydrostatic models, and physically the velocity fields in stellar atmospheres have no relation to the observer's viewing angle (i.e. the $\mu$ angle, as the calculations of radiation transfer are carried from the observer's point of view). In the subsequent sections, we show the impact and biases in the abundances associated with these approximations of 1D hydrostatic equilibrium and/or LTE.
\subsection{{3D} NLTE solar yttrium abundance}

\begin{figure}[ht]
\includegraphics[width=1\columnwidth]{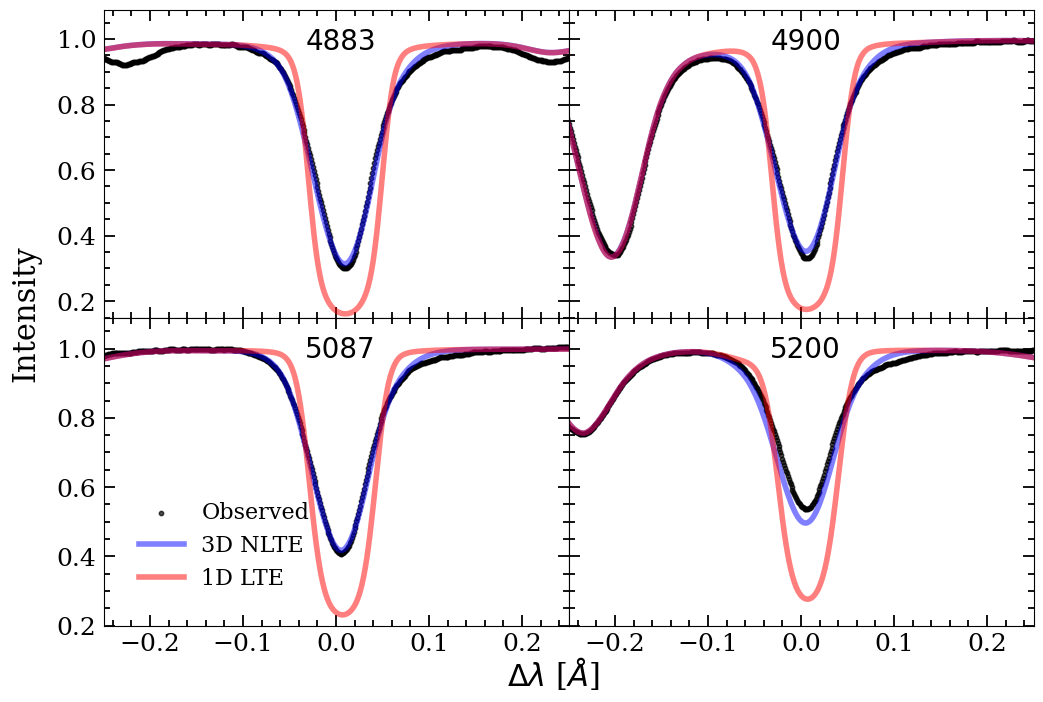}
\caption{{Comparison of the observed Y II line profiles from the solar disc-centre ($\mu = 1.00$) intensity atlas (black) with the 3D NLTE (blue) and 1D LTE (red) model profiles with the instrumental broadening and a constant microturbulence of $\vmic = 1$ km/s for 1D models. A(Y) = 2.3 dex is used for all models for clear comparison.}}
 \label{fig:y_lines_examples_1.0}
\end{figure}

\begin{figure}[ht]
\includegraphics[width=1\columnwidth]{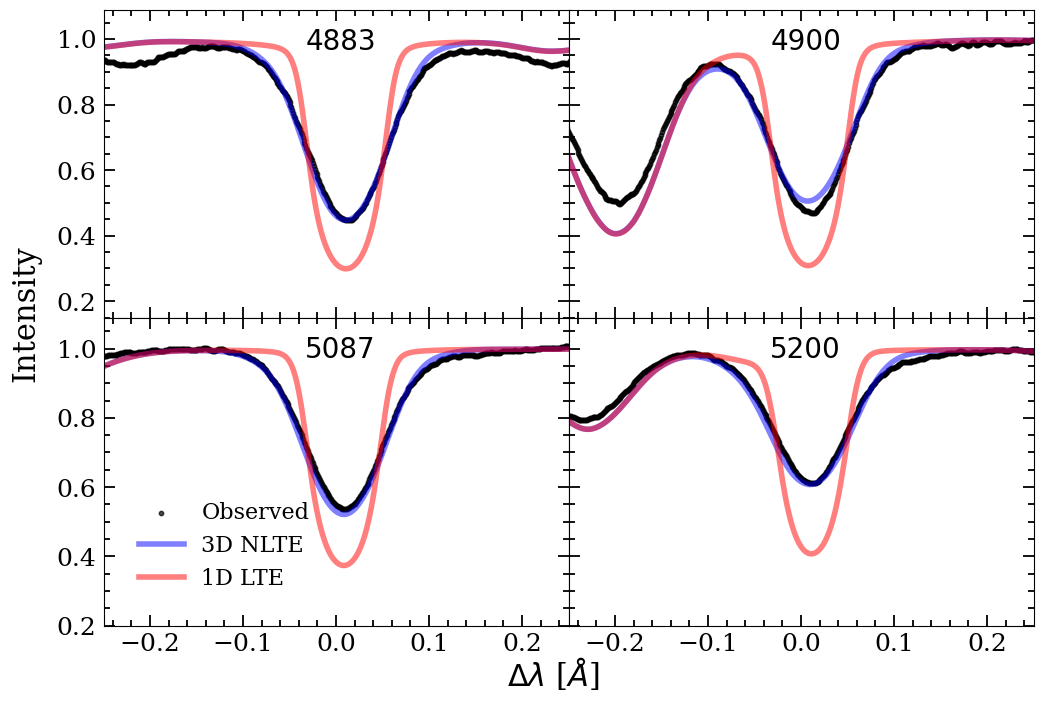}
\caption{Same as Fig. \ref{fig:y_lines_examples_1.0} but for the limb, $\mu = 0.20$.}
 \label{fig:y_lines_examples_0.2}
\end{figure}

\begin{figure}
\includegraphics[width=1\columnwidth]{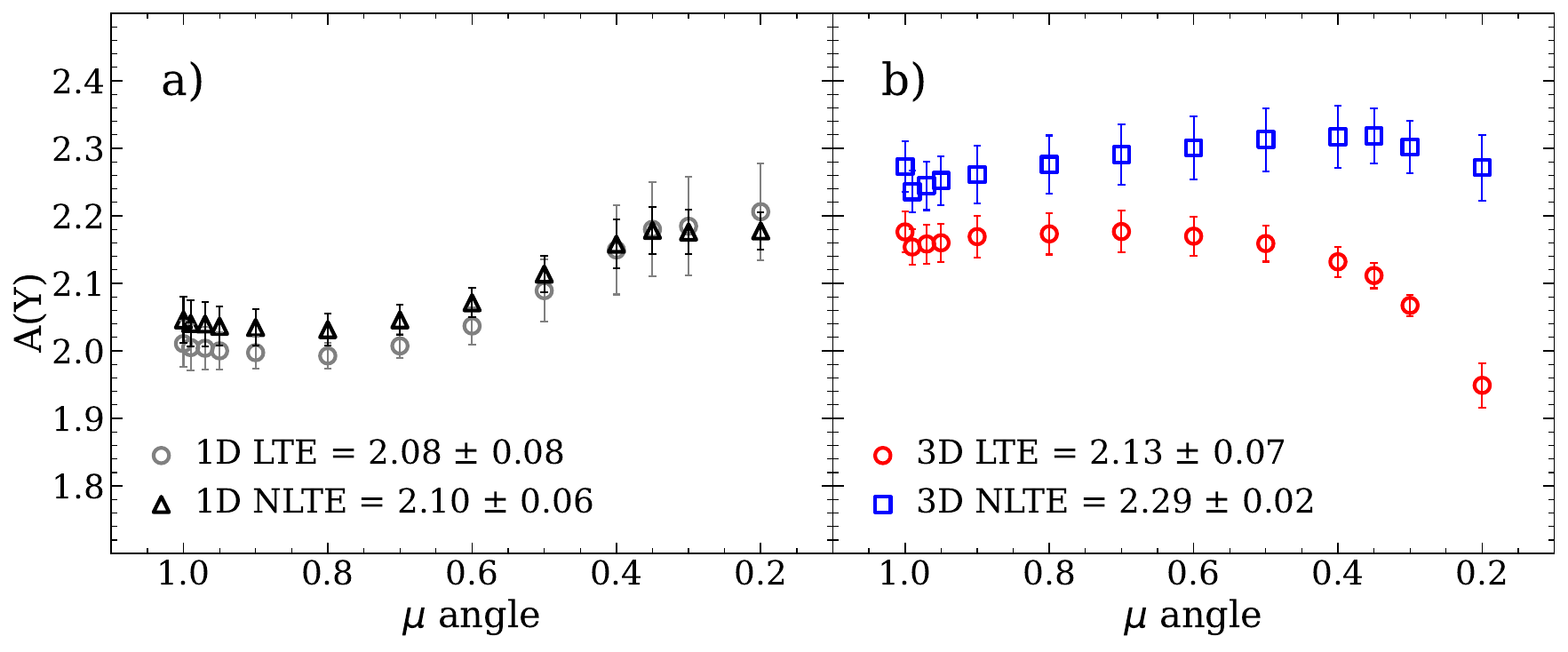}
\caption{{Abundances of Y determined from the IAG high-resolution solar observations taken at different viewing angles ($\mu$) using 1D LTE, 1D NLTE, 3D LTE, and 3D NLTE models. The results obtained using 1D LTE and 1D NLTE line formation models are shown in the left panel (a). The results obtained using 3D LTE and 3D NLTE models are shown in the right panel (b). The average solar A(Y) value and its standard deviation are provided in the figure inset for each model. See the main text for further details.}}
 \label{fig:y_fit_4883}
\end{figure}

Figures \ref{fig:y_lines_examples_1.0} and \ref{fig:y_lines_examples_0.2} show the resulting 1D LTE and 3D NLTE line profiles of Y II lines for two $\mu$ angles, $\mu = 1.00$ and $0.20$, respectively. The observed lines tend to be weaker and broader at the limb compared to the disc centre. {We highlight the limitations of 1D hydrostatic models by avoiding artificial broadening in synthetic line profiles, except for necessary instrumental broadening} and $\Vmic$ for the 1D model. Clearly, 1D hydrostatic models are unable to describe the shapes of the observed data and their angular variation, requiring angle-dependent macro-turbulence to compensate for the apparent increasing line broadening. In contrast, 3D NLTE profiles successfully reproduce the observed solar spectra in black without the need for any artificial broadening. This evidence is not new, but this is the first time that detailed CLV spectroscopic datasets for Y II are tested against 1D and 3D solar models in LTE and NLTE, adding to the body of evidence suggesting that 3D NLTE modelling is critical for careful CLV work, as known for other species, such as O \citep{Bergemann2021, Pietrow2023}.

The final results of 1D LTE, 1D NLTE, 3D LTE, and 3D NLTE abundance analyses of Y are presented in Fig. \ref{fig:y_fit_4883}. Here, we include a constant $\vmic = 1$ km/s and $\vmac$ for 1D models. Interestingly, the behaviour of recovered abundances at different angles differs between different modelling approaches. 1D LTE and 1D NLTE modelling approaches lead to a progressively increasing bias, suggesting a systematically increasing abundance towards the limb, which is clearly unphysical. As such, the disc-centre abundance retrieval in 1D yields an  A(Y) of $\approx 2.0$ dex, whereas the limb values range from A(Y) $\approx 2.1$ to $2.2$ dex, depending on the Y II line. A very similar systematic bias, albeit in the opposite direction ---abundances systematically lower at the limb--- is present in the 3D LTE results. Specifically, the 3D LTE disc-centre values are nearly $0.2$ dex higher compared to the values obtained at the limb. This curvature cannot be corrected by adjusting macroturbulence $\vmac$ (unless in the case of strong fine-tuning), but could possibly be mitigated by introducing a $\mu$-angle-dependent $\vmic$. However, this is not the purpose of the present study, as we are primarily aiming to testing our 3D NLTE models and we simply present the classical 1D and/or LTE results for comparison.

Remarkably, 3D NLTE modelling allows us to achieve significantly more consistent abundances across the solar disc for all diagnostic Y II lines (Fig. \ref{fig:y_fit_4883}), thereby also yielding a substantially reduced abundance scatter. A weak rising trend is still present as a function of $\mu$ angle; however, it may be that the residual discrepancy is due to the blends or continuum placement. The $4900$ and $5200$ \AA~ lines in particular show a systematic bias; however, both features are also visibly affected by strong blends in the blue wing and by a weaker blend in the red wing in both Fig. \ref{fig:y_lines_examples_1.0} and Fig. \ref{fig:y_lines_examples_0.2}. The 5200 \AA~line is also weaker by at least 20-30\% compared to the other three lines. We therefore chose to exclude this line from the final abundance determination. However, we note that the Y abundance derived from this line, {A(Y) $= 2.24 \pm 0.03_{\textrm{stat}} \pm 0.07_{\textrm{syst}}$} dex, is entirely consistent with those of the other diagnostic Y II features.

\begin{table}
\setlength{\tabcolsep}{3pt}
\begin{minipage}{\linewidth}
\renewcommand{\footnoterule}{} 
\caption{Derived weighted average Y abundances with uncertainty coming {only from variations of abundances at different $\mu$ angles}. We do not use the line 5200 $\AA$ in the final abundance determination, but we note that its 3D NLTE value lies close to those of the other lines.}
\label{tab:y_abund}     
\begin{center}
\begin{tabular}{c c c c c c}
\noalign{\smallskip}\hline\noalign{\smallskip}    
\AA~~ & 1D LTE & 1D NLTE & 3D LTE & 3D NLTE  \\
\noalign{\smallskip}\hline\noalign{\smallskip}  

4883 & 2.10 $\pm$ 0.13 & 2.08 $\pm$ 0.08 & 2.13 $\pm$ 0.08 & 2.30 $\pm$ 0.03 \\
4900 & 2.10 $\pm$ 0.09 & 2.13 $\pm$ 0.07 & 2.14 $\pm$ 0.09 & 2.35 $\pm$ 0.03 \\
5087 & 2.09 $\pm$ 0.07 & 2.11 $\pm$ 0.05 & 2.15 $\pm$ 0.07 & 2.27 $\pm$ 0.02 \\
5200 & 2.02 $\pm$ 0.04 & 2.07 $\pm$ 0.04 & 2.09 $\pm$ 0.04 & 2.24 $\pm$ 0.03 \\

\noalign{\smallskip}\hline\noalign{\smallskip}
\end{tabular}
\end{center}
\end{minipage}
\end{table} 

The values of our investigated Y II lines are provided in Table \ref{tab:y_abund}. Our final abundance estimate, as a weighted average over all $\mu$ angles for the first three diagnostic lines, is as follows: {A(Y)$_{\textrm{3D NLTE}}$~=~$2.30 \pm 0.03_{\textrm{stat}} \pm 0.07_{\textrm{syst}}$. For comparison, A(Y)$_{\textrm{1D LTE}}$~=~$2.10 \pm 0.10_{\textrm{stat}} \pm 0.07_{\textrm{syst}}$, A(Y)$_{\textrm{1D NLTE}}$~=~$2.11 \pm 0.07_{\textrm{stat}} \pm 0.07_{\textrm{syst}}$, and A(Y)$_{\textrm{3D LTE}}$~=~$2.14 \pm 0.08_{\textrm{stat}} \pm 0.07_{\textrm{syst}}$}. The error is computed as described in Sect. \ref{subsec:error_analysis}. The weights for calculating both the average and the standard deviation are based on the actual distances between consecutive data points in terms of the angle $\mu$. This approach ensures that each segment along the $\mu$ axis contributes equally to the calculations, compensating for the uneven spacing of the data points. {While it is plausible that the standard deviation between the lines already accounts for the systematic errors, it is difficult to devise an experiment to confirm this independently. We decided to provide both, making it a conservative estimate of our error.} Overall, the differences are expected given the properties of the statistical equilibrium of Y and the NLTE effects on Y~II lines \citep{Storm2023}. The 1D NLTE abundances are slightly higher compared to 1D LTE, except for the line 4883. In contrast, all of the 3D NLTE values are systematically higher compared to 3D LTE, which is interesting. As opposed to the diagnostic lines of Ba II \citep{Gallagher2020}, which are extremely strong in the solar spectrum ---and therefore the behaviour is set by the behaviour of the source function---, the Y~II lines are weak and are particularly sensitive to the variation of line opacity in 3D NLTE. Line formation in inhomogeneous 3D atmospheres is extremely complex, owing to the correlated motions and temperature and density fluctuations. It is therefore remarkable that 3D modelling yields consistent abundance values across the solar disc, without the need for any fine-tuning of models. Our results also indicate that it is not possible to generalise results obtained for one chemical element, as is sometimes done in the literature. One such example can be found in \citet{Asplund2009, Asplund2021}, where the authors adopt an NLTE correction for vanadium based on atoms with similar atomic numbers.

Our study is the first 3D NLTE analysis of Y. However, we can compare our estimates with those of \citet{Grevesse2015}, who carried out 1D and 3D LTE abundance analyses of Y for the solar disc-centre. Firstly, our error is larger because it also includes $f$-value uncertainty. However, it is not clear whether it was done in \citet{Grevesse2015}. Secondly, we confirm their findings that 3D LTE disc-centre abundances are $\sim 0.05$ to $0.10$ dex higher than those suggested by 1D LTE, especially for the Y II lines in common. However, owing to the differences in the choice of the diagnostic lines and their $\log gf$ values (see our Sect. \ref{subsec:diag_lines}) and Table 1 in \citet{Grevesse2015}, our 3D LTE abundances are not exactly identical, namely {A(Y)$_{\textrm{3D LTE}}$~=~$2.14 \pm 0.08_{\textrm{stat}} \pm 0.07_{\textrm{syst}}$ (this work)} but A(Y)$_{\textrm{3D LTE}}$ = $2.21$ \citep{Grevesse2015}, although they are still consistent within the combined uncertainties of both values. However, after careful inspection of the line selection of \citet{Grevesse2015}, we found that most of their lines are not suitable for a reliable solar analysis, owing to significant blending and/or problems with the continuum.

For the sake of completeness, we also derived the 3D NLTE Y abundance from the solar flux data. We obtained A(Y) = $2.26 \pm 0.07$ and A(Y) = $2.30 \pm 0.07$ for the IAG data \citep{Reiners2016} and the KPNO data, respectively. These results are fully consistent with our values obtained from the disc-resolved intensities. The differences between the IAG and KPNO results are primarily due to the normalisation of the spectra. Our values are slightly higher than the abundance of Y in CI chondrites, of  A(Y)$_{\textrm{meteoric}} = 2.17 \pm 0.04$ \citep{Lodders2009}. However, we note that the latter depends on the choice of the solar Si abundance used to normalise the cosmochemical abundance scale to the photospheric scale, and the minor differences are not of concern in this work.
\subsection{{3D} NLTE solar Eu abundance}

\begin{table}
\begin{minipage}{\linewidth}
\renewcommand{\footnoterule}{} 
\setlength{\tabcolsep}{1.9pt}
\caption{Derived A(Eu) abundances for IAG and KPNO flux and disc-centre spectra. The average error for all of the measurements is $\sim 0.06$ dex, where the dominant source of error is the uncertainty on the $f$-value \citep[based on that reported in][]{Lawler2001}.}
\label{tab:eu_abund}     
\begin{center}
\begin{tabular}{l | c c c c}
\noalign{\smallskip}\hline\noalign{\smallskip}    
Data & 1D LTE & 1D NLTE & 3D LTE & 3D NLTE \\
\noalign{\smallskip}\hline\noalign{\smallskip}   

IAG flux         & 0.57 & 0.58 & 0.57 & 0.55 \\
IAG $\mu=1.0$    & 0.58 & 0.58 & 0.59 & 0.57 \\
KPNO flux        & 0.59 & 0.60 & 0.59 & 0.57 \\
KPNO $\mu=1.0$   & 0.60 & 0.60 & 0.60 & 0.57 \\
\noalign{\smallskip}\hline\noalign{\smallskip}
\end{tabular}
\end{center}
\end{minipage}
\end{table}

\begin{figure}
\centering
\includegraphics[width=1\columnwidth]{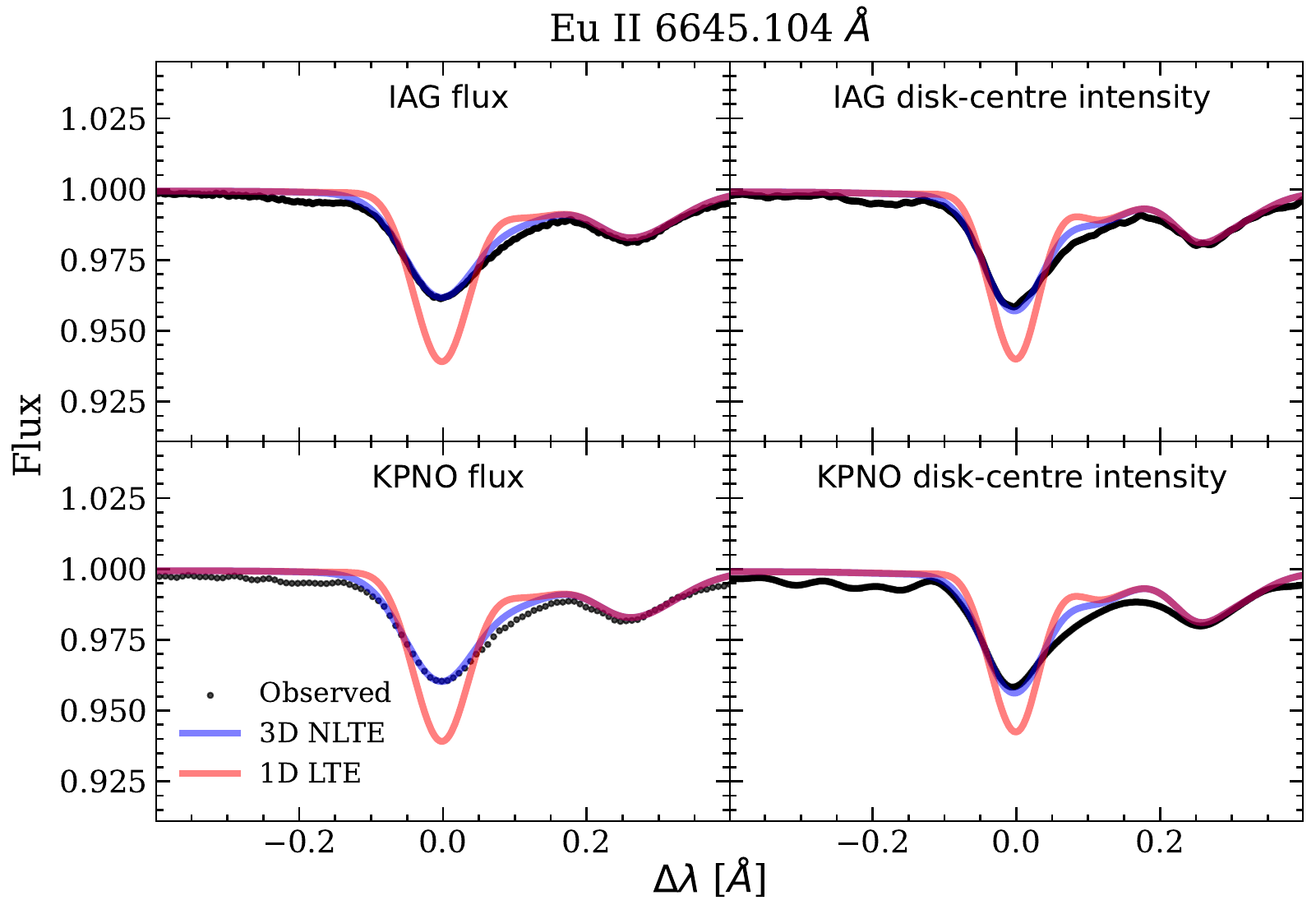}
\caption{Comparison of the observed Eu II 6645 line profile in IAG and KPNO spectra (in black) with the 3D NLTE (in blue) and 1D LTE (in red) models of the closest abundance within 0.01 dex only with instrumental broadening and rotational broadening for the flux, and constant microturbulence $\vmic = 1$ km/s for 1D models.}
 \label{fig:eu_fit_example}
\end{figure}

Figure \ref{fig:eu_fit_example} shows the 3D NLTE, 1D LTE, and observed profiles of the  diagnostic Eu II line at 6645 \AA. The observations refer to the IAG and KPNO data, and we show both the flux and disc-centre intensity fits. The KPNO data were slightly renormalised to improve the mismatch in the continuum level. It is rather unfortunate that this key diagnostic of Eu is significantly affected by line blending, which makes it difficult to derive a robust Eu abundance from this feature. This was also the case in previous studies; specifically \citet{Mashonkina2000} and \citet{Mucciarelli2008}, who ascribed the blending to the lines of Cr~I and Si~I. In the Gaia-ESO line list, we also find other lines in the relevant wavelength range; however, the contribution appears to be smaller. As in these studies, we include HFS and isotopic splitting and carefully fit the models to the observed data, taking the blends into account. For Si \citep{Bergemann2013} and Cr \citep{Bergemann2010b}, 3D LTE calculations are sufficient, because of negligibly small NLTE effects for the relevant lines.

Our resulting 3D NLTE solar abundance of Eu is {A(Eu) = $0.57 \pm 0.01_{\textrm{stat}} \pm 0.06_{\textrm{syst}}$} dex. Other values, computed in 1D LTE, 1D NLTE, and 3D LTE, are also provided in Table \ref{tab:eu_abund}. {1D} LTE yields a slightly larger abundance. The NLTE model of Eu results in a slightly positive correction in 1D and a negative one in 3D. The 3D LTE values are $\sim 0.01$ dex higher compared to the 1D LTE results, which is fully consistent with the findings of \citet{Mucciarelli2008}.

In terms of NLTE effects, we find a similar, small,  positive 1D NLTE correction to that found by \citet{Mashonkina2000}. However, interestingly, we note that the sign of the NLTE correction for the 6645 \AA~line would be flipped if we were to use the Drawin recipe for Eu$+$H collisions, which is ---at a first glance--- unexpected. Unfortunately, the authors did not quote the number of radiative transitions included in their model, and their NLTE model was not published. Therefore, a one-to-one comparison and a detailed comparative analysis cannot be performed. Nonetheless, useful insights can be gained from an exploratory analysis performed by reducing the complexity of our model. As such, in an attempt to reproduce the results of \citet{Mashonkina2000}, we performed a series of tests, resorting to a different statistical equilibrium code called DETAIL \citep{Butler1985}, reducing the size of the model atom, eliminating radiative transitions (bound--bound and bound--free), scaling the photo-ionisation cross-sections, and substituting the quantum-mechanical collisional data with the Drawin recipe and applying scaling factors to the latter as done by \citet{Mashonkina2000}.

The results of our tests suggest that the property that has a major influence on the statistical equilibrium of Eu is the physical completeness of the atomic model. The study by \citet{Mashonkina2000} relied on a model atom with 32 Eu II states that is closed by the ground state of Eu III. For comparison, our Eu model comprises 662 energy levels, with 236 of these being Eu II. We find that we can reproduce the results of \citet{Mashonkina2000}, that is positive NLTE abundance corrections with the Drawin recipe for H$+$Eu with a scaling coefficient of one-third, by reducing the model atom to 44 Eu II states ---which is roughly consistent with their work (closed by the level \Eu{e}{7}{S}{o}{})---, and removing a large fraction of strong radiative transitions connecting \Eu{a}{9}{D}{o}{} with upper terms, specifically, \Eu{x}{7}{P}{o}{}, \Eu{x}{9}{P}{o}{}, \Eu{y}{9}{P}{o}{}, and \Eu{z}{7}{P}{o}{} with wavelengths primarily in the near-UV and blue ranges. In contrast, even eliminating the Eu~I ion or photo-ionising reactions entirely has no effect on the statistical equilibrium of Eu~II. Likewise, we find that scaling the Drawin recipe only has an appreciable effect when a very small atomic model is used. We therefore conclude that the differences with the earlier work by \citet{Mashonkina2000} are primarily explained by the more comprehensive model atom in our work.

Our final 3D NLTE solar abundance is {A(Eu) = $0.57 \pm 0.01_{\textrm{stat}} \pm 0.06_{\textrm{syst}}$} dex. This estimate is higher than the estimate by \citet{Grevesse2015} and \citet{Asplund2021} of A(Eu) $ = 0.52$ dex. However, we note that in these latter works, 3D LTE and 1D NLTE corrections were applied separately, by applying a 3D LTE correction to the 1D LTE results from \citet{Lawler2001} and subsequently applying a $+0.03$ dex 1D NLTE correction for the flux rather than intensity from \citet{Mashonkina2000}. Also, our error is slightly higher, because we include the uncertainty of atomic data ($f$-values) explicitly, which dominates the total error budget. Unfortunately, again, it is not clear whether this was done in \citet{Grevesse2015}. Our solar 3D NLTE abundance of Eu is fully consistent and is derived directly by 3D NLTE spectrum synthesis, without any corrections involving other models.
\section{Conclusions}
\label{sec:discussion}
In this paper, we focus on a 3D NLTE analysis of the solar abundances of two key s- and r-process elements, Y and Eu, respectively. We developed a model atom of Eu using quantum-mechanical data on inelastic processes in collisions of europium with hydrogen, and updated our comprehensive model of Y \citep{Storm2023} with the quantum-mechanical data on inelastic processes in collisions of yttrium with hydrogen. We also explore the NLTE effects in both ions and compare our results with the literature. We performed our abundance analysis using recent high-quality solar data from the IAG atlases \citep{Reiners2016, Ellwarth2023}, both fluxes and spatially resolved intensities, and we  complement them with the analysis of the KPNO FTS datasets from \citet{Kurucz1984}.

{Our findings suggest that the NLTE effects in the solar diagnostic lines of Y II and Eu II are of a different nature, confirming results in the literature. For Eu II, we find very small and positive NLTE effects on the key diagnostic line at 6645 \AA, similar to the findings of \citet{Mashonkina2000}. We also show that in full 3D radiative-transfer calculations, the abundances are only slightly lower compared to 1D NLTE. The differences between 3D LTE and 1D (MARCS) LTE Eu abundances are also very small and positive, which is consistent with \citet{Mucciarelli2008}. For Y II, despite the addition of quantum-mechanical collisional data, we find no significant changes to the 1D NLTE solar abundance compared to \citet{Storm2023}, which is not unexpected, as that study also demonstrated that Y II lines are not particularly sensitive to the treatment of inelastic H collisional processes. However, 3D effects in the Y II lines are significant, and increase the solar Y abundance by $\sim 0.2$ dex compared to 1D LTE. The differences between 3D LTE and 1D LTE Y abundances are positive and range from $+0.03$ to $+0.07$ dex, overall qualitatively supporting the findings of \citet{Grevesse2015}.}

{As an additional test of our models, we performed a detailed analysis of the centre-to-limb variation of the diagnostic lines. For Eu II, this test was not successful, because of strong blending that prohibited accurate diagnostics at large angles close to the limb. Despite our best efforts, we still lack knowledge about the blends in the corresponding wavelength region. For Y II on the other hand, we obtain consistent 3D NLTE Y abundance at all $\mu$ angles, with minor deviations at the limb. We also show that LTE leads to a systematic bias, with abundances obtained from the disc-centre observations being significantly lower (with the 1D MARCS model) or higher (with the 3D RHD model) than that derived from limb observations.}

Our final 3D NLTE Y and Eu solar abundances are {A(Y)$_{\textrm{3D NLTE}}$ = $2.30 \pm 0.03_{\textrm{stat}} \pm 0.07_{\textrm{syst}}$} dex and {A(Eu) = $0.57 \pm 0.01_{\textrm{stat}} \pm 0.06_{\textrm{syst}}$} dex. The total errors include the errors of the $f$-values and, for Y, the systematic error component assessed by modelling the abundance of this element across the entire solar disc. While our results are different from those of previous work, we show that the the differences are fully expected and are caused by  the new NLTE atomic models, atomic data, and self-consistent 3D NLTE radiative transfer modelling. Our study is the first self-consistent 3D NLTE analysis of these elements, while previous work relied {on} 1D LTE \citep{Lawler2001}, 1D NLTE \citep{Mashonkina2000}, and 1D LTE co-added with 3D LTE and 1D NLTE corrections \citep{Grevesse2015, Asplund2021}. The overall good agreement between the observed and modelled line profiles of Eu and Y, as well as consistent abundances for the flux and spatially resolved intensity observations are encouraging and confirm that the 3D NLTE approach offers the best way to perform accurate and precise chemical abundance studies of the Sun and by implication of all other stars. {All stellar abundance analyses are tailored to the solar abundance values. We therefore anticipate that our new 3D NLTE measurements will serve as a more accurate benchmark for different stellar surveys, such as WEAVE, 4MOST, and SDSS-V. Our group is currently working on grid-based 3D NLTE methodologies to enable comprehensive modelling for large stellar samples, and this study represents an important reference for all subsequent work building on physically advanced 3D NLTE models.}

\begin{acknowledgements}
MB is supported through the Lise Meitner grant from the Max Planck Society. This project has received funding from the European Research Council (ERC) under the European Unions Horizon 2020 research and innovation programme (Grant agreement No. 949173). We acknowledge support from the DAAD grant Project No.: 57654415. PP and PQ are, respectively, Research Associate and Research Director of the Belgian Fund for Scientific Research F.R.S. - FNRS.  P.S.B. acknowledges support from the Swedish Research Council through an individual project grant with contract no. 2020-03404, and support through the project “Probing charge- and mass-transfer reactions on the atomic level,” from the Knut and Alice Wallenberg Foundation (2018.0028). SAY and AKB gratefully acknowledge Project No. 22-23-01181 support (RScF).

We sincerely thank the referee for the positive feedback and comments that improved the readability of the paper.
\end{acknowledgements}

\bibliographystyle{aa}
\bibliography{references, paul}

\appendix

\label{lastpage}
\end{document}